\documentclass[UTF-8]{sn-jnl}
\usepackage{caption}
\usepackage{graphicx, subfig}
\usepackage{verbatim}
\usepackage{graphicx}
\usepackage{bbding}
\usepackage{amsmath}
\usepackage{amssymb}
\usepackage{amsfonts}
\usepackage{mathrsfs}
\usepackage{mathrsfs, amssymb}
\usepackage{booktabs}
\usepackage[perpage,symbol*]{footmisc}
\usepackage[misc]{ifsym}

\def\vbeta{\mbox{\boldmath$\beta$}}
\def\vtheta{\mbox{\boldmath$\theta$}}

\def\vOmega{\mbox{\boldmath$\Omega$}}

\def\veta{\mbox{\boldmath$\eta$}}

\def\vSigma{\mbox{\boldmath$\Sigma$}}
\def\a{\mbox{\boldmath$a$}}

\def\e{\mbox{\boldmath$e$}}

\def\E{\mbox{\boldmath$E$}}

\def\I{\mbox{\boldmath$I$}}
\def\b{\mbox{\boldmath$b$}}
\def\u{\mbox{\boldmath$u$}}

\def\V{\mbox{\boldmath$V$}}
\def\x{\mbox{\boldmath$x$}}

\def\Z{\mbox{\boldmath$Z$}}

\def\0{\mbox{\boldmath$0$}}

\def\cov{\mbox{cov}}

\def\drow{\stackrel{d}{\longrightarrow}}
\def\prow{\stackrel{p}{\longrightarrow}}

\jyear{2021}%

\theoremstyle{thmstyleone}%
\theoremstyle{thmstyletwo}%

\theoremstyle{thmstylethree}%

\begin{document}

\title[Optimal subsampling algorithm for CQR with distributed data]{Optimal subsampling algorithm for composite quantile regression with distributed data}


\author[1]{\fnm{Xiaohui} \sur{Yuan}}\email{yuanxh@ccut.edu.cn}

\author[1]{\fnm{Shiting} \sur{Zhou}}\email{zhoushiting1999@outlook.com}
\equalcont{These authors contributed equally to this work.}

\author*[1]{\fnm{Yue} \sur{Wang}}\email{wangyueccut@gmail.com}
\equalcont{These authors contributed equally to this work.}

\affil*[1]{\orgdiv{School of Mathematics and Statistics}, \orgname{Changchun University of Technology}, \city{Changchun}, \postcode{130012}, \state{Jilin}, \country{China}}


\abstract{ For massive data  stored at multiple machines, we propose a distributed subsampling procedure for the composite quantile regression. By establishing the consistency and asymptotic normality of the composite quantile regression estimator from a general subsampling algorithm, we derive the optimal subsampling probabilities and the optimal allocation sizes under the L-optimality criteria.  A two-step algorithm to approximate the optimal subsampling procedure is developed. The proposed methods are illustrated through numerical experiments on simulated and real datasets.}


\maketitle

\section{Introduction}\label{sec1}
With the rapid development of science and technology, extremely large datasets are ubiquitous and lays heavy burden on storage and computation facilities.
Many efforts have been made to deal with these challenge. There are three main directions from the view of statistical applications: divide-and-conquer, online updating, and subsampling. Among them, subsampling has been found to be useful for reducing computational burden and extracting information from massive data.

The idea of subsampling was first proposed by Jones (1956)\cite{jones1956}. A key tactic of subsampling methods is to specify nonuniform sampling probabilities to include more informative data points with higher probabilities. For example, the leverage score-based subsampling  in Ma et al. (2015)\cite{Ma2015}, the information based optimal subdata selection in Wang et al. (2019)\cite{Wang2019},  and the optimal subsampling method under the A-optimality criterion in Wang et al. (2018)\cite{wang2018}. Recently, Fang et al. (2021)\cite{fang2021} applied subsampling to a weak-signal-assisted procedure for variable selection and statistical inference. Ai et al. (2021)\cite{ai2021} studied the optimal subsampling method for generalized linear models under the A-optimality criterion.
Shao et al. (2022)\cite{ShaoL2022} employed the optimal subsampling method to ordinary quantile regression.

Due to the large scale and fast arrival speed of data stream, massive data are often partitioned across multiple servers. For example, Walmart stores produce a large number of data sets from different locations around the world, which need to be processed. However, it is difficult to transmit these datasets to a central location. For these datasets, it is common to analyze them on multiple machines. Qiu et al. (2020)\cite{qiu2020} constructed a data stream classification model based on distributed processing.  Sun et al. (2021)\cite{sun2021} proposed a data mining scheme for edge computing based on distributed integration strategy. Zhang and Wang (2021)\cite{zhand2021} proposed a distributed subdata selection method for big data linear regression model. Zuo et al. (2021)\cite{zuo2021} proposed a distributed subsampling procedure  for the logistic regression. Yu et al. (2022)\cite{yu2022} derived a optimal distributed Poisson subsampling procedure for the maximum quasi-likelihood estimators with massive data.

In the paper, we investigate the optimal distributed subsampling for composite quantile regression (CQR; Zou and Yuan (2008)\cite{zou2008}) in massive data.
In a linear model, composite quantile regression can uniformly estimate the regression coefficients under heavy tail error. Moreover, since the asymptotic variance of the composite quantile regression estimate does not depend on the moment of the error distribution, the CQR estimator is robust. The CQR method is widely used in many fields. For massive data, Jiang et al. (2018)\cite{jiang2018}  proposed a divide-and-conquer CQR method.  Jin and Zhao (2021)\cite{jin2021} proposed a divide-and-conquer CQR neural network method.  Wang et al. (2021)\cite{wang2021} proposed a distributed CQR method for the massive data. Shao and Wang (2022)\cite{ShaoY2021} and Yuan et al. (2022)\cite{yuan2022} developed the subsampling for composite quantile regression. To the best of our knowledge, there is almost no work on random subsampling for composite quantile regression with distributed data.

Based on the above motivation, we investigate the optimal subsampling for the composite quantile regression in massive data when the datasets are stored at different sites. We propose a distributed subsampling method in the context of CQR, and then study the optimal subsampling technology for data in each machine.
The main advantages of our method are as follows: First, we establish the convergence rate of the subsample-based estimator, which ensures the consistency of our proposed method. Second, it avoids the impact of different intercept items in data sets stored at different sites. Third, the computational speed of our subsampling method is much faster than the full data approach.

The rest of this article is organized as follows. In Section 2, we propose the distributed subsampling algorithm based on composite quantile regression. The asymptotic properties of estimators based on subsamples are also established. We present a subsampling strategy with optimal subsampling probability and optimal allocation size. The simulation studies are given in Section 3. In Section 4, we study the real data sets. The content of the article is summarized in Section 5. All proofs are given in the Appendix.

\section{Methods}\label{sec2}

\subsection{Model and notation}
Consider the following linear model
\begin{eqnarray}\label{Logisic}
  y_{ik} &=& \x_{ik}^\textsf{T}\vbeta_0+\varepsilon_{ik},\ i=1,\cdots,n_{k},\ k=1,\cdots,K,
\end{eqnarray}
where $\x_{ik}$ denotes a $p$-dimensional covariate vector, $\vbeta_{0}=(\beta_1,\cdots,\beta_{p})^\textsf{T} \in \Theta$ is a $p$-dimensional vector of regression coefficients, $n_{k}$ is the sample size of the $k$th dataset, $n=\sum_{k=1}^{K} n_{k}$ is the total sample size, and $K$ is the number of distributed datasets. Assume that the random error $\varepsilon_{ik}$ has cumulative distribution function $F(\cdot)$ and probability density function $f(\cdot)$.

Let $M$ be the composite level of composite quantile regression, which does not depend on the sample size $n$. Given $M$, let $\tau_{m},m=1,\cdots,M$ be the specified quantile levels such that $\tau_{1}<\cdots <\tau_{M}$. Write $\vtheta_0=(\theta_{01},\cdots,\theta_{0(p+M)})^\textsf{T}=(\vbeta_{0}^\textsf{T},\b_{0}^\textsf{T})^\textsf{T}$ and $\b_0=(b_{01},\cdots,b_{0M})^\textsf{T}$, where $b_{0m}=\inf\{u: F(u)\geq\tau_m\}$ for $m=1,\cdots,M$. In this paper, we assume that $\x_{ik}$'s are nonrandom and are interested in inferences about the unknown $\vtheta_0$ from the observed dataset
$$D_n=\{D_{kn}=\{(\x_{ik}^\textsf{T},y_{ik}),\ i=1\cdots,n\},\ k=1,\cdots, K\}.$$
For $\tau\in(0,1)$, $u\in R^{p}$, let $\rho_{\tau}(u)=u\{\tau -I(u < 0)\}$ be the check loss function for the $\tau$-th quantile level. The CQR estimator of $\vtheta$ based on the full dataset $D_{n}$ is given by
\begin{eqnarray}\label{distX}
  \hat{\vtheta}_{F}=(\hat{\vbeta}_{F}^\textsf{T},\hat{\b}_{F}^\textsf{T})^\textsf{T}=\mathop{\arg\min}\limits_{\text{\vbeta},\text{\b}} \sum_{k=1}^{K}\sum_{i=1}^{n_{k}}\sum_{m=1}^{M} \rho_{\tau_{m}} (y_{ik}-b_{m}-\x_{ik}^\textsf{T}\vbeta),
\end{eqnarray}
Our aim is to construct a subsample-based estimator, which can be used to effectively approximate the full data estimator $\hat{\vtheta}_{F}$.

\subsection{ Subsampling algorithm and asymptotic properties}
In this subsection, we propose a distributed subsampling algorithm to approximate the $\hat{\vtheta}_{F}$. First we propose a subsampling method in Algorithm 1, which can reasonably select a subsample from distributed data.

\hspace*{\fill} \\
\noindent
\textbf{Algorithm 1} Distributed Subsampling Algorithm£º
\begin{itemize}
  \item  Sampling:
Assign subsampling probabilities $\{\pi_{ik}\}_{i=1}^{n_{k}}$ for the $k$th dataset $D_{k}=\{(y_{ik},\x_{ik}),i=1,\cdots,n_{k}\}$ with $\sum_{i=1}^{n_{k}}\pi_{ik}=1$, where $k=1,\cdots,K$. Given total sampling size $r$, draw a random subsample of size $r_{k}$ with replacement from $D_{k}$ according to $\{\pi_{ik}\}_{i=1}^{n_{k}}$, where $\{r_{k}\}_{k=1}^{K}$ are allocation sizes with $\sum_{k=1}^{K}r_{k}=r$. For $i=1,\cdots,n_{k}$ and $k=1,\cdots,K$, we denote the corresponding responses, covariates, and subsampling probabilities as $y_{ik}^{\ast},\x_{ik}^{\ast}$ and $\pi_{ik}^{\ast}$, respectively.
  \item  Estimation:
Based on the subsamples $\{(y_{ik}^{\ast},\x_{ik}^{\ast},\pi_{ik}^{\ast}),i=1,\cdots,r_{k}\}_{k=1}^{K}$, and calculate the estimate $\tilde{\vtheta}_{s}=(\tilde{\vbeta}_{s},\tilde{\b}_{s})=\arg\min_{\text{\vtheta}} Q^{\ast}(\vtheta)$, where
\begin{eqnarray*}
  Q^{\ast}(\vtheta) &=&\frac{1}{n} \sum_{k=1}^{K} \frac{r}{r_{k}}\sum_{i=1}^{r_{k}}\sum_{m=1}^{M} \frac{\rho_{\tau_{m}} (y_{ik}^{\ast}-\vbeta^\textsf{T}\x_{ik}^{\ast}-b_{m})}{\pi_{ik}^{\ast}}.
\end{eqnarray*}
\end{itemize}

To establish asymptotic properties of the subsample-based estimator $\tilde{\vtheta}_{s}$, we need the following assumptions:

\noindent

\textbf{(A.1)} Assume that $f(t)$ is continuous with respect to $t$ and $0 < f(b_{0m}) < +\infty$ for $1 \leq m \leq M$. Let $\tilde{\x}_{ik,m}=(\x_{ik}^\textsf{T},\e_{m}^\textsf{T})^\textsf{T}$, where $\e_{m}$ denotes a $M \times 1$ vector, which has a one only in its $m$th coordinate and is zero elsewhere.
Define
\begin{eqnarray}
  \E_{n} &=&\frac{1}{n}\sum_{k=1}^{K}\sum_{i=1}^{n_{k}}\sum_{m=1}^{M} f(b_{0m})\tilde{\x}_{ik,m}(\tilde{\x}_{ik,m})^\textsf{T}.
\end{eqnarray}
Assume that there exist positive definite matrices $\E$, such that
\begin{eqnarray*}
 \E_n \longrightarrow \E,\ \ \mbox{and}\ \ \max_{1\leq k\leq K, 1\leq i\leq n_{k}} \|\x_{ik}\| &=& o(n^{1/2}).
\end{eqnarray*}

\noindent

\textbf{(A.2)} Assume that, for $k=1,\cdots,K$.
\begin{eqnarray}
\max_{1\leq k\leq K, 1\leq i\leq n_{k}} \frac{\|\x_{ik}\|+1 }{r_{k}\pi_{ik}}=o_p\left(\frac{n}{r^{1/2}}\right).
\end{eqnarray}
Define
\begin{eqnarray}
  \V_{\pi} &=& \frac{1}{n^{2}}\sum_{k=1}^{K}\frac{r}{r_{k}} \sum_{i=1}^{n_{k}}\frac{1}{\pi_{ik}}\left[\sum_{m=1}^{M}\{I(\varepsilon_{ik} < b_{0m})-\tau_{m}\}\tilde{\x}_{ik,m}\right]^{\otimes2},
\end{eqnarray}
where for a vector $\a$, $\a^{\otimes 2} =\a \a^\textsf{T}$. Assume that there exist positive definite matrices $\V$ such that
\begin{eqnarray*}
  \V_{\pi} \prow \V,
\end{eqnarray*}
where $\prow$ means convergence in probability.

\noindent

\textbf{Theorem 1}. If Assumptions (A.1) and (A.2) hold, conditional on $D_{n}$, as $n \rightarrow \infty$ and $r \rightarrow \infty$, if $r/n =o(1)$, then we have
\begin{eqnarray}
  \vSigma^{-1/2} \sqrt{r} (\tilde{\vtheta}_{s} -  \vtheta_{0}) \drow N(\0,\I),
\end{eqnarray}
where $\drow$ denotes convergence in distribution, $\vSigma=\E_{n}^{-1} \V_{\pi} \E_{n}^{-1}$.

\subsection{Optimal subsampling strategy}\label{sec3}

Given $r$, we specify the subsampling probablities $\{\pi_{ik}\}_{i=1}^{n_{k}}$, and the allocation sizes $\{r_{k}\}_{k=1}^{K}$ in Algorithm 1. A naive choice is the uniform subsampling strategy with
$\{\pi_{ik}=1/n_{k}\}_{i=1}^{n_{k}}$ and $\{r_{k}=[rn_{k}/n]\}_{k=1}^{K}$, where $[\cdot]$ denotes the rounding operation. However, this uniform subsampling method is not optimal. As suggested by Wang et al. (2018)\cite{wang2018}, we adopted the nonuniform subsampling strategy to determine the optimal allocation sizes and optimal subsampling probabilities by minimizing the trace of $\vSigma$ in Theorem 1.

Since $\vSigma=\E_{n}^{-1} \V_{\pi} \E_{n}^{-1}$, the optimal allocation sizes and subsampling probabilities require the calculation of $\E_{n}$, which depend on the unknown density function $f(\cdot)$. Following Wang and Ma (2021)\cite{wangM2021}, we derive optimal subsampling probabilities under the L-optimality criterion. Note that $\E_{n}$ and $\V_{\pi}$ are nonnegative definite. Simple matrix algebra yields that $tr(\vSigma)=tr(\V_{\pi} \E_{n}^{-2})=tr(\E_{n}^{-2})tr(\V_{\pi})$. $\vSigma$ depends on $r_{k}$ and $\pi_{ik}$ only through $\V_{\pi}$, and $\E_{n}$ is free of $r_{k}$ and $\pi_{ik}$. Hence, we suggest to determine the optimal allocation sizes and optimal subsampling probabilites by directly minimizing $tr(\V_{\pi})$ rather than $tr(\vSigma)$, which can effectively speed up our subsampling algorithm.

\noindent

\textbf{Theorem 2}.  If $r_{k}$ and $\pi_{ik}$, $i=1,\cdots,n_{k}$, $k=1,\cdots,K$, are chosen as
\begin{eqnarray}\label{pilopt}
\pi_{ik}^{Lopt}= \pi_{ik}^{Lopt}(\vtheta_0)= \frac{\parallel\sum_{m=1}^{M}\{\tau_{m}-I(\varepsilon_{ik} < b_{0m})\}\tilde{\x}_{ik,m}\parallel}{\sum_{i=1}^{n_{k}} \parallel\sum_{m=1}^{M}\{\tau_{m}-I(\varepsilon_{ik} < b_{0m})\}\tilde{\x}_{ik,m}\parallel},
\label{pik}
\end{eqnarray}
and
\begin{eqnarray}
  r_{k}^{Lopt} &=& r\frac{\sum_{i=1}^{n_{k}} \parallel\sum_{m=1}^{M}\{\tau_{m}-I(\varepsilon_{ik} < b_{0m})\}\tilde{\x}_{ik,m}\parallel}{\sum_{k=1}^{K} \sum_{i=1}^{n_{k}} \parallel\sum_{m=1}^{M}\{\tau_{m}-I(\varepsilon_{ik} < b_{0m})\}\tilde{\x}_{ik,m}\parallel},
  \label{rk}
\end{eqnarray}
then $tr(\V_\pi)/n$ attains its minimum.

\subsection{Two-step algorithm}\label{sec4}

Note that the optimal subsampling probabilities and allocation sizes depend depends on  $\varepsilon_{ik}=y_{ik}-\x^T_{ik} \vbeta_0$ and $b_{0m}$, $m=1,\cdots, M$.
The L-optimal weight result is not directly implementable. To deal with this problem, we use a pilot estimator $\tilde{\vtheta}$ to replace $\vtheta_0$. In the following, we propose a two-step subsampling procedure in Algorithm 2.

\hspace*{\fill} \\
\noindent
\textbf{Algorithm 2} Two-Step Algorithm£º
\begin{itemize}
  \item  Step 1:
Given $r_{0}$, we run Algorithm 1 with subsampling size $r_{k}=[r_{0}\frac{n_{k}}{n}]$ to obtain a pilot estimator $\tilde{\vtheta}$, using $\pi_{ik}=1/n_{k}$, where $[\cdot]$ denotes the rounding operation. Replace $\vtheta_0$ with $\tilde{\vtheta}_{0}$ in (\ref{pik}) and (\ref{rk}) to get the allocation sizes $r_{k}(\tilde{\vtheta})$ and subsampling probabilities $\pi_{ik}(\tilde{\vtheta})$, for $i = 1,\cdots, n_{k}$ and $k = 1, \cdots, K$, respectively.
  \item  Step 2:
Based on $\{r_{k}(\tilde{\vtheta})\}_{k=1}^{K}$ and $\{\pi_{ik}(\tilde{\vtheta})\}_{i=1}^{n_{k}}$ in Step 1, we can select a subsample $\{(y_{ik}^{\ast},\x_{ik}^{\ast},\pi_{ik}^{\ast}):i=1,\cdots,r_{k}\}_{k=1}^{K}$ from the full data $D_{n}$. Minimizes the following weighted function
\begin{eqnarray*}
  Q^{\ast}(\vtheta) &=& \sum_{k=1}^{K} \frac{r}{r_{k}(\tilde{\vtheta})}\sum_{i=1}^{r_{k}(\tilde{\text{\vtheta}})}\sum_{m=1}^{M} \frac{\rho_{\tau_{m}} (y_{ik}^{\ast}- \vbeta^\textsf{T} \x_{ik}^{\ast} -b_{m})}{\pi_{ik}^{\ast}},
\end{eqnarray*}
to get a two-step subsample estimate $\hat{\vtheta}_{Lopt}$, where $\hat{\vtheta}_{Lopt}=(\hat{\vbeta}_{Lopt},\hat{\b}_{Lopt})=\arg\min Q^{\ast}(\vtheta)$.
\end{itemize}

For the subsample-based estimator $\hat{\vtheta}_{Lopt}$ in Algorithm 2, we give its asymptotic distribution in the following theorem.

\noindent

\textbf{Theorem 3}. If Assumptions (A.1) and (A.2) hold, then as $r_{0}\rightarrow\infty$, $r\rightarrow\infty$, and $n\rightarrow\infty$, then we have
\begin{eqnarray}
  \vSigma^{-1/2} \sqrt{r} (\hat{\vtheta}_{Lopt}-\vtheta_{0}) \drow N(\0,\I),
\end{eqnarray}
where $\drow$ denotes convergence in distribution, $\vSigma=\E_{n}^{-1} \V_{\pi} \E_{n}^{-1}$. Here
\begin{eqnarray}
  \V_{\pi} &=& \frac{1}{n^{2}}\sum_{k=1}^{K}\frac{r}{r_{k}^{Lopt}} \sum_{i=1}^{n_{k}}\frac{1}{\pi_{ik}^{Lopt}}\left[\sum_{m=1}^{M}\{I(\varepsilon_{ik} < b_{0m})-\tau_{m}\}\tilde{\x}_{ik,m}\right]^{\otimes2},
\end{eqnarray}
where
$$\pi_{ik}^{Lopt}=\frac{\parallel\sum_{m=1}^{M}\{\tau_{m}-I(\varepsilon_{ik} < b_{0m})\}\tilde{\x}_{ik,m}\parallel}{\sum_{i=1}^{n_{k}} \parallel\sum_{m=1}^{M}\{\tau_{m}-I(\varepsilon_{ik}< b_{0m})\}\tilde{\x}_{ik,m}\parallel},$$
and
$$r_{k}^{Lopt} = r\frac{\sum_{i=1}^{n_{k}} \parallel\sum_{m=1}^{M}\{\tau_{m}-I(\varepsilon_{ik} < b_{0m})\}\tilde{\x}_{ik,m}\parallel}{\sum_{k=1}^{K} \sum_{i=1}^{n_{k}} \parallel\sum_{m=1}^{M}\{\tau_{m}-I(\varepsilon_{ik} < b_{0m})\}\tilde{\x}_{ik,m}\parallel}.$$

For the statistical inference about $\vtheta_0$, to avoid estimating $f(b_{0m})$, we propose the following iterative sampling procedure.

Firstly, using $\{\pi_{ik}^{Lopt}(\tilde{\vtheta})\}_{i=1}^{n_{k}}$ proposed in Algorithm 2, we sample with replacement to obtain $B$ subsamples, $\{(y_{ik}^{\ast,j},\x_{ik}^{\ast,j},\pi_{ik}^{\ast,j}), i=1,\cdots,r_{k}^{Lopt}(\tilde{\vtheta}), k=1,\cdots,K\}$ for $j=1,\cdots,B$. Next, we calculate the $j$th estimate of $\vtheta_0$ through
\begin{eqnarray*}
      \hat{\vtheta}_{Lopt,j} &=& (\hat{\vbeta}_{Lopt,j},\hat{\b}_{Lopt,j})\\
      &=& \arg\min_{\text{\vtheta}} \sum_{k=1}^{K} \frac{r}{r_{k}^{Lopt}(\tilde{\vtheta})}\sum_{i=1}^{r_{k}^{Lopt}(\tilde{\text{\vtheta}})} \sum_{m=1}^{M}\frac{\rho_{\tau_{m}} (y_{ik}^{\ast,j}- \vbeta^\textsf{T} \x_{ik}^{\ast,j} -b_{m})}{\pi_{ik}^{\ast,j}}.
\end{eqnarray*}
The combined estimate can be obtained by
\begin{eqnarray}\label{zzf}
       \hat{\vtheta}_L=(\hat{\vbeta}_{L}^\textsf{T},\hat{\b}_{L}^\textsf{T})^\textsf{T} = \frac{1}{B}\sum_{j=1}^B \hat{\vtheta}_{Lopt,j}
\end{eqnarray}
and its variance-covariance matrix $\vOmega=\cov(\hat{\vtheta}_L)$ can be estimated by
\begin{eqnarray}\label{omg}
      \hat{\vOmega} &=& \frac{1}{r_{ef} B(B-1)}\sum_{j=1}^B (\hat{\vtheta}_{Lopt,j}-\hat{\vtheta}_L)^{\otimes 2},
\end{eqnarray}
where $r_{ef}$ is the effective subsample size ratio (Wang \& Ma, 2021\cite{wangM2021}) given by
\begin{eqnarray*}
       r_{ef} &=&\frac{1}{K}\sum_{k=1}^{K}\left(1- \frac{r_{k}B-1}{2}\sum_{i=1}^{n_{k}} \{\pi_{ik}^{Lopt}(\tilde{\vtheta})\}^2\right).
\end{eqnarray*}

From Theorem 3, for any fixed $B$, the conditional distribution of $\sqrt{rB}(\hat{\vtheta}_{L}-\vtheta_0)$ satisfies
\begin{eqnarray*}
 \{\E_n^{-1} \V_{\pi} \E_n^{-1}\}^{-1/2} \sqrt{rB}(\hat{\vtheta}_{L}- \vtheta_0) & \drow & N(\0,\I).
\end{eqnarray*}

The distribution of $\hat{\vtheta}_{Lopt}$ can be approximated by the empirical distribution of $\{\tilde{\vtheta}_{Lopt,j}\}_{j=1}^B$. For $s=1,\cdots,p+K$, the $100\times(1-\alpha)$\% confidence interval of $\theta_{0s}$ can be approximated by $[\hat{\theta}_{L,s}-\hat{\omega}_{ss}^{1/2}z_{1-\alpha/2},\hat{\theta}_{L,s}+\hat{\omega}_{ss}^{1/2}z_{1-\alpha/2}]$, where $\hat{\theta}_{L,s}$ is the $s$th element of $\hat{\vtheta}_L$, $\hat{\omega}_{ss}$ is the $(s,s)$th element of $\hat{\vOmega}$ and $z_{1-\alpha/2}$ is the $1-\alpha/2$ quantile of the standard normal distribution.

\section{Numerical studies}\label{sec5}

In this section,  we conduct a simulation study to evaluate the performances of the proposed optimal
subsampling algorithm.  Simulations were performed on a laptop running Window 10 with an Intel i7 processor and 16 GB memory. Full data are generated from the model
\begin{eqnarray*}
  y_{ik} &=& \x^\textsf{T}_{ik} \vbeta_{0} +\varepsilon_{ik},\ i=1,\cdots,n_k,\ k=1,\cdots,K,
\end{eqnarray*}
with the true parameter $\vbeta_{0}=(1,1,1,1,1)^{T}$. We consider the following four cases for the error term $\varepsilon$:
(1) the standard normal distribution, $N(0,1)$;
(2) the mixture normal distribution, $0.5N(0,1) + 0.5N(0,9)$;
(3) the Student¡¯s t distribution with three degrees of freedom, $t(3)$;
(4) the standard Cauchy distribution, Cauchy(0,1).

We consider the following four cases for the covariate $\x$:

Case I: $\x_{ik} \sim N(\0,\vSigma)$, where $\vSigma=(0.5^{\mid s-t\mid})_{s,t}$.

Case II: $\x_{ik} \sim N(\0,\vSigma)$, where $\vSigma=(0.5^{I(s\neq t)})_{s,t}$.

Case III: $\x_{ik} \sim t_{3}(\0,\vSigma)$ with three degrees of freedom and $\vSigma=(0.5^{\mid s-t\mid})_{s,t}$.

Case IV: Set $K = 5$, $\x_{i1}\sim N_{5}(\0, \I)$, $\x_{i2}\sim N_{5}(\0, \vSigma_{1})$, $\x_{i3}\sim N_{5}(\0, \vSigma_{2})$, $\x_{i4}\sim t_{3}(\0,\vSigma_{1})$ and  $\x_{i5}\sim t_{5}(\0,\vSigma_{1})$, where $\vSigma_{1}= (0.5^{\mid s-t\mid})_{s,t}$, $\vSigma_{2}= (0.5^{I(s\neq t)})_{s,t}$.

Note that in Cases I-III, the covariate distributions are identical for all distributed datasets. In Case IV, the covariates have different distributions for distributed datasets.

All the simulation are based on 1000 replications. We set the sample size of each datasets as $\{n_{k}=[nu_{k}/\sum_{k=1}^{K}u_{k}]\}_{k=1}^{K}$, where $[\cdot]$ denotes the rounding operation, $u_{k}$ are generated from the uniform distribution over (1, 2) with $K = 5$ and 10, respectively. We use the quantile levels $\tau_{m}=m/16,m=1,\cdots,15$ for the composite quantile regression.

In Tables 1, we report the simulation results on subsample-based estimator for $\beta_{1}$ (other $\beta_{i}$'s are similar and omitted) with $K=5$ and $K=10$ respectively, including the estimated bias (Bias) and the standard deviation (SD) of the estimates where $r_{0} = 200, n = 10^{6}$  in Case I.  The bias and SDs of the proposed subsample estimate   for  Case IV with $n=10^6$ and $n=10^7$ are presented in Tabel 2.  The subsample sizes $r = 200, 400, 600, 800$ and $1000$, respectively. It can be seen from the results that the subsample-based estimator is unbiased. The performance of our estimator becomes better as $r$ increases, which confirms the theoretical result on consistency of the subsampling methods.

For comparison, we consider the uniform subsampling method (Uniform) with $\pi_{ik}=\frac{1}{n_{k}}$, and $r_{k}=[rn_{k}/n]$  for $i=1,\cdots,n_{k}$ and $k=1,\cdots,K$. We calculate empirical mean square error (MSE) of uniform subsampling estimator (Unif) and our optimal subsampling estimator (Lopt) based on 1000 repetitions of the simulation. Figures 1 and 2 present the MSEs of each method for Case I with $K = 5$ and $K = 10$, where $n = 10^{6}$.  Figures 3 presents the  MSEs of the subsampling estimator for Case IV with $n=10^6$, $n=10^7$ and $\varepsilon \sim N(0,1)$.  From the above results, we can see that the MSEs of our method (Lopt) are much smaller than those of Uniform subsampling method (Unif).  The results indicate that our method also works well with heterogeneous covariates, i.e., the covariates can have different distributions in different data blocks.

In the following, we evaluate the computational efficiency of our two-step subsampling algorithm. The mechanism of data generation is the same as the above mentioned situation. For fair comparison, we count the CPU time with one core based on the mean calculation time of 1000 repetitions of each subsample-based method. In Table 3, we report the results for Case I and the normal error with $n = 10^{6}, K = 5, r_{0} = 200$ and  different $r$, respectively. The computing time for the full data method is also given in the last row. Note that the uniform subsampling requires the least computing time, because its subampling probabilities $\pi_{ik}=\frac{1}{n_{k}}$, and allocation sizes $r_{k}=[rn_{k}/n]$, do not take time to compute. Our subsampling algorithm has great computation advantage over the full data method. To further investigate the computational gain of the subsampling approach, we increase the dimension $p$ to $30$ with the true parameter $\vbeta_{0}=(0.5,\cdots,0.5)^\textsf{T}$. Table 4 presents the computing time for   Case I and normal error  with $r_{0} = 200, r = 1000, K = 5, n = 10^{4},10^{5},10^{6}$ and $10^{7}$, respectively. It is clear that both subsampling methods take significantly less computing times than the full data approach.

To investigate  the performance of $\hat{\vOmega}$ in (\ref{omg}), we compare the empirical mean square error (EMSE, $s^{-1}\sum_{s=1}^{1000}\parallel \hat{\vbeta}_{L}^{s}-\vbeta_{0}\parallel^{2}$) and the average estimated mean square error(AMSE) of $\hat{\vbeta}_L$ in (\ref{zzf}) with different   $B$. In Tables 5, we report the average length of the confidence intervals  and 95\% coverage probabilities (CP) of our subsample-based estimator for $\beta_{1}$ (other $\beta_{i}$'s are similar and omitted) with $n = 10^{6}, r = 1000$ and $K = 5$. Figures 4-7 present the EMSEs and AMSEs of $\hat{\vbeta}_L$. For all cases, the AMSEs are very close to the EMSEs, and the EMSEs and AMSEs become smaller as $B$ increases.

\section{A real data example}\label{sec6}

In this section, we apply our method to the USA airline data, which are publicly available at http://stat-computing.org/datastore/2009/the-data.html. The data include detailed information on the arrivals and departures of all commercial flights in the USA from 1987 to 2008, and they are stored in 22 separate files ($K=22$). The raw dataset is as large as 10 GB on a hard drive. We use the composite  regression to model the relationship between the arrival delay time, $y$, and three covariate variables: $x_1$, weekend/weekday status (binary; 1 if departure occurred during the weekend, 0 otherwise), $x_2$, the departure delay time and $x_3$,  the distance. Since the $y$, $x_2$ and $x_3$ in the data set are on different scales, we normalize them first. In addition, we drop the NA values in the dataset and we have $n = 115,257,291$ observations with completed information on $y$ and $\x$. Table 6 shows the cleaned data.

We  use the quantile levels $\tau_{m}=m/16,m=1,\cdots,15$ for the composite quantile regression.  For comparison, the full-data estimate of the regression parameters is given by ${\hat{\vbeta}}_{F}=(-0.0451,0.9179,-0.0248)^\textsf{T}$. The proposed point estimate $\hat{\vbeta}_L$ and corresponding confident intervals with different $r$ and $B$ are presented in Table 7. It can be seen from Table 7 that the subsample estimator $\hat{\vbeta}_L$ is close to ${\hat{\vbeta}}_{F}$.
In Figure 8, we present the MSEs of both subsampling methods based on 1000 subsamples with $r=200$, 400, 600, 800 and 1000, respectively. The MSEs of the the optimal subsampling estimator  are smaller than those of the uniform subsampling estimator.

\section{Conclusion}\label{sec7}

We have studied the statistical properties of a subsampling algorithm for the composite quantile regression model with distributed massive data. We derived the optimal subsampling probabilities and optimal allocation sizes. The asymptotic properties of the subsample estimator were established. Some simulations and a real data example were provided to check the performance of our method.

\begin{appendices}

\section*{Appendix}

\noindent \textbf{Proof of Theorem 1}

\noindent
Define
\begin{eqnarray*}
A_r^*(\u) &=& \frac{1}{n}\sum_{k=1}^K \frac{r}{r_{k}} \sum_{i=1}^{r_{k}} \sum_{m=1}^M\frac{1}{\pi^*_{ik}} A^*_{ik,m}(\u),
\end{eqnarray*}
where
$A^*_{ik,m}(\u) = \rho_{\tau_m}( \varepsilon_{ik}^*-b_{0m}- \u^\textsf{T} \tilde{\x}_{ik,m}^* /\sqrt{r})-\rho_{\tau_m} ( \varepsilon_{ik}^*-b_{0m})$,  $\tilde{\x}_{ik,m}^*=(\x_{ik}^{*\textsf{T}}, \e_{m}^\textsf{T})^\textsf{T}$, and $\varepsilon_{ik}^*=y_{ik}^*-\vbeta^\textsf{T}_0 \x_{ik}^*$,  $i=1,\cdots,r_{k}$. Since $A_r^*(\u)$ is a convex function of $\u$, its minimizer is $ \sqrt{r}( \tilde{\vtheta}_s-\vtheta_0)$, we can focus on  $A_r^*(\u)$ when evaluating the properties of $ \sqrt{r}( \tilde{\vtheta}_s-\vtheta_0)$.

Let $\psi_\tau (u)=\tau-I(u<0)$.  By Knight's identity (Knight, 1998),
\begin{eqnarray*}
 \rho_{\tau}(u-v)-\rho_{\tau}(u)&=& -v\psi_\tau (u)+\int_0^v \{I(u\leq s)-I(u\leq 0)\}ds,
\end{eqnarray*}
we can rewrite $A^*_{ik,m}(\u)$ as
\begin{eqnarray*}
 A^*_{ik,m}(\u)     &=& \rho_{\tau_m} ( \varepsilon_{ik}^*-b_{0m}-  \u^\textsf{T}\tilde{\x}_{ik,m}^* /\sqrt{r})-\rho_{\tau_m}( \varepsilon_{ik}^*-b_{0m})\\
                  &=& - \frac{1}{\sqrt{r}} \u^\textsf{T}\tilde{\x}_{ik,m}^* \{\tau_m- I(\varepsilon_{ik}^*-b_{0m}<0 )\}\\
                  &&+ \int_0^{ \u^\textsf{T}\tilde{\x}_{ik,m}^*/\sqrt{r}} \{I(\varepsilon_{ik}^*- b_{0m}\leq s)-I(\varepsilon_{ik}^*-b_{0m}\leq0)\}ds.
\end{eqnarray*}
Thus, we have
\begin{eqnarray}
          & & A_r^*(\u)  \nonumber\\
          &=& -\u^\textsf{T}\frac{1}{\sqrt{r}} \frac{1}{n} \sum_{k=1}^K \frac{r}{r_{k}} \sum_{m=1}^M\sum_{i=1}^{r_{k}} \frac{1}{\pi^*_{ik}} \{\tau_m- I(\varepsilon_{ik}^*-b_{0m}<0 )\}\tilde{\x}_{ik,m}^* \nonumber\\
          & & + \frac{1}{n}\sum_{k=1}^K \frac{r}{r_{k}}  \sum_{m=1}^M\sum_{i=1}^{r_{k}}  \frac{1}{\pi^*_{ik}}\int_0^{ \u^\textsf{T}\tilde{\x}_{ik,m} /\sqrt{r}} \{I(\varepsilon_{ik}^*-b_{0m}\leq s)-I(\varepsilon_{ik}^*-b_{0m}\leq 0)\}ds\nonumber\\
          &=& \u^\textsf{T} \Z_{r}^*  +A^*_{2r}(\u),\label{Andecom0}
\end{eqnarray}
where
\begin{eqnarray*}
 & & \Z_{r}^*= - \frac{1}{\sqrt{r}} \frac{1}{n}\sum_{k=1}^K \frac{r}{r_{k}} \sum_{m=1}^M\sum_{i=1}^{r_{k}} \frac{1}{ \pi^*_{ik}} \{\tau_m- I(\varepsilon_{ik}^*-b_{0m}<0 )\}\tilde{\x}_{ik,m}^*,\\
 & & A^*_{2r}(\u)= \frac{1}{n}\sum_{k=1}^K \frac{r}{r_{k}} \sum_{i=1}^{r_{k}} \frac{1}{\pi^*_{ik}} A^*_{k,i}(\u), \\
 & & A^*_{k,i}(\u)= \sum_{m=1}^M \int_0^{\u^\textsf{T}\tilde{\x}_{ik,m}^* /\sqrt{r}} \{I(\varepsilon_{ik}^*-b_{0m}\leq s)-I(\varepsilon_{ik}^*-b_{0m}\leq0)\}ds.
\end{eqnarray*}
Firstly, we prove the asymptotic normality of $\Z_{r}^* $. Denote
\begin{eqnarray*}
\veta_{ik}^* &=& -\frac{r}{r_{k}n\pi^*_{ik}} \sum_{m=1}^M \{\tau_m- I(\varepsilon_{ik}^*-b_{0m}<0)\}\tilde{\x}_{ik,m}^*,
\end{eqnarray*}
then $\Z_{r}^*$ can be written as $\Z_{r}^*= \frac{1}{\sqrt{r}} \sum_{k=1}^K  \sum_{i=1}^{r_{k}} \veta_{ik}^*$. Direct calculation yields
\begin{eqnarray*}
 E(\veta_{ik}^*\mid D_n) &=&- \frac{r}{r_{k}n} \sum_{i=1}^{n_{k}} \sum_{m=1}^M \{\tau_m-I(\varepsilon_{ik}-b_{0m}<0 )\}\tilde{\x}_{ik,m}=O_p \left(\frac{r n_k^{-1/2}}{r_{k}n}\right), \\
 cov(\veta_{ik}^*\mid D_n) &=&  E\{(\veta_{ik}^*)^{\otimes 2}\mid D_n\}-\{E(\veta_{ik}^*\mid D_n)\}^{\otimes 2} \\
  &=&   \sum_{i=1}^{n_{k}} \frac{r^2}{r_{k}^2 n^2\pi_{ik}} \biggr\{\sum_{m=1}^M  [\tau_m-I(\varepsilon_{ik}-b_{0m}<0)] \tilde{\x}_{ik,m}\biggr\} ^{\otimes 2}- \{E(\veta_{ik}^*\mid D_n)\}^{\otimes 2} \\
  &=&   \sum_{i=1}^{n_{k}} \frac{r^2}{r_{k}^2 n^2\pi_{ik}} \biggr\{\sum_{m=1}^M  [\tau_m-I(\varepsilon_{ik}-b_{0m}<0)] \tilde{\x}_{ik,m}\biggr\} ^{\otimes 2} - o_p(1).
\end{eqnarray*}
It is easy to verify that
\begin{eqnarray*}
&& E\{E(\veta_{ik}^*\mid D_n)\}= 0,  \\
&&cov\{E(\veta_{ik}^*\mid D_n)\}=  \frac{r^2}{r_{k}^2 n^2}\sum_{i=1}^{n_{k}} cov\left\{\sum_{m=1}^M \left[ \tau_m-I(\varepsilon_{ik}<b_{0m})\right]\tilde{\x}_{ik,m}\right\}.
\end{eqnarray*}
Denote the $(s,t)$ th element of $ cov\{E(\veta_{ik}^*\mid D_n)\}$ as $\sigma_{st}$. Using the Cauchy inequality, it is easy to obtain $$\mid\sigma_{st}\mid \leq \sqrt{\sigma_{ss}}\sqrt{\sigma_{tt}}\leq \frac{r^2}{r_{k}^2 n^2}\sum_{i=1}^{n_{k}} M(\|\x_i\|^2+1)=O_p\left(\frac{r^2 n_{k}}{r_{k}^2 n^2}\right).$$ By Assumption 1 and Chebyshev's inequality, $$ E(\veta_{ik}^* \mid D_n) =O_p\left(\frac{r n_{k}^{1/2}}{r_{k} n}\right).$$

Under the conditional distribution given $D_n$, we check Lindeberg's conditions (Theorem 2.27 of van der Vaart, 1998). Specifically, for $\epsilon>0$, we want to prove that
\begin{eqnarray}
&&\sum_{k=1}^{K}\sum_{i=1}^{r_{k}} E\{\|r^{-1/2}  \veta_{ik}^* \|^2 I(\|\veta_{ik}^*\|>\sqrt{r}\epsilon )  \mid D_n\}=o_{p}(1).
 \label{Ex2}
\end{eqnarray}
Note that
 \begin{eqnarray}
&&\sum_{k=1}^{K}\sum_{i=1}^{r_{k}} E\{\|r^{-1/2}  \veta_{ik}^* \|^2 I(\|\veta_{ik}^*\|>\sqrt{r}\epsilon )  \mid D_n\}
\nonumber\\
&=& \sum_{k=1}^{K}\sum_{i=1}^{r_{k}} E\biggr\{\biggr\|\frac{r^{1/2}}{r_{k} n \pi_{ik}^*}  \sum_{m=1}^M \tilde{\x}_{ik,m}^*\{\tau_m-I(\varepsilon_{ik}-b_{0m}<0 )\} \biggr\|^2
\nonumber\\
&& \times I\biggr(\biggr\|\frac{r^{-1/2}}{ r_{k} n \pi_{ik}^* \epsilon }  \sum_{m=1}^M \tilde{\x}_{ik,m}^* \{\tau_m-I(\varepsilon_{ik}-b_{0m}<0)\} \biggr\|>1 \biggr)  \biggr\lvert D_n \biggr\}
\nonumber\\
&=& \sum_{k=1}^{K} \sum_{i=1}^{n_{k}} \frac{r}{ r_{k} n^2 \pi_{ik}} \biggr\| \sum_{m=1}^M \{\tau_m-I(\varepsilon_{ik}-b_{0m}<0)\}\tilde{\x}_{ik,m} \biggr\|^2
\nonumber\\
&& \times I\biggr(\frac{r^{1/2}}{ r_{k} n \pi_{ik} \epsilon} \biggr\| \sum_{m=1}^M \{\tau_m-I(\varepsilon_{ik}-b_{0m}<0)\}  \tilde{\x}_{ik,m} \biggr\| >1 \biggr).
 \label{Expeta}
\end{eqnarray}
By Assumption (A.2), $$\max_{1\leq k\leq K}\max_{1\leq i\leq n_{k}} \frac{\|\x_{ik}\|+1 }{r_{k}\pi_{ik}}=o_p\left(\frac{n}{r^{1/2}}\right),$$ $$M^2\sum_{k=1}^{K}\sum_{i=1}^{n_{k}}\frac{(1+ \| \x_{ik}\|)^2}{n^2\pi_{ik}}=O_p(1),$$
the right hand side of (\ref{Expeta}) satisfies
 \begin{eqnarray}
&& \sum_{k=1}^{K}\sum_{i=1}^{n_{k}}\frac{r}{ r_{k} n^2 \pi_{ik}} \biggr\| \sum_{m=1}^M \{\tau_m-I(\varepsilon_{ik}<b_{0m})\}\tilde{\x}_{ik,m} \biggr\|^2
\nonumber\\
&&\times I\biggr(\frac{r^{1/2}}{ r_{k} n \pi_{ik} \epsilon} \biggr\| \sum_{m=1}^M \{\tau_m-I(\varepsilon_{ik}<b_{0m})\}  \tilde{\x}_{ik,m} \biggr\| >1 \biggr)
\nonumber\\
&\leq & M^2 \sum_{k=1}^{K}\sum_{i=1}^{n}\frac{r}{ r_{k} n^2 \pi_{ik}} (1+ \| \x_{ik}\|)^2  I\biggr(\frac{M(1+ \|\x_{ik}\|)r^{1/2}}{ r_{k} n \pi_{ik} \epsilon}>1 \biggr)
\nonumber\\
&\leq& I\biggr(\max_{1\leq k\leq K}\max_{1\leq i\leq n_{k}} \frac{\|\x_{ik}\|+1 }{r_{k}\pi_{ik}}>\frac{n\epsilon}{r^{1/2} M} \biggr)\nonumber\\
&&\times M^2\sum_{k=1}^{K}\sum_{i=1}^{n_{k}}\frac{r(1+ \| \x_{ik}\|)^2}{r_{k} n^2 \pi_{ik}}\nonumber\\
&=& o_{p}(1).
 \end{eqnarray}
Thus, the Lindeberg's conditions hold with probability approaching one.

Note that $\veta^*_{ik}$, $i=1,\cdots,r_{k}$, are independent and identically distributed with mean $E(\veta_{ik}^*  \mid D_n)$ and the covariance $cov(\veta_{ik}^* \mid D_n)$ when given $D_n$. Based on this result, as $r, n \rightarrow \infty$, we get
\begin{eqnarray*}
\V_{\pi}^{-1/2}\{ \Z_{r}^*- \sqrt{r} \sum_{k=1}^{K}E(\veta_{ik}^*\mid D_n)\} &\drow& N(\0,\I).
\end{eqnarray*}

Since $\sqrt{r} \sum_{k=1}^{K}E(\veta_{ik}^*\mid D_n)=O_p\left(\frac{r^{1/2}}{n^{1/2}}\sum_{k=1}^{K}\frac{r n_{k}^{1/2}}{r_{k} n^{1/2}}\right)=o_p(1)$, it is easy to verify that
\begin{eqnarray}\label{Zndistri}
\V_{\pi}^{-1/2} \Z_{r}^*  & \drow & N(\0,\I).
\end{eqnarray}

Next, we prove that
\begin{eqnarray*}
A^*_{2r}(\u)&=&  \frac{1}{2}\u^\textsf{T}\E\u+o_{p}(1).
\end{eqnarray*}
Write the conditional expectation of $A^*_{2r}(\u)$ as
\begin{eqnarray}\label{EA2nk}
&&E\{A^*_{2r}(\u) \mid D_n\} \nonumber\\
 &=& \frac{r}{n}\sum_{k=1}^K \sum_{i=1}^{n_{k}} E\{A_{k,i}(\u)\}+\frac{r}{n}\sum_{k=1}^K\sum_{i=1}^{n_{k}} [ A_{k,i}(\u)-E\{A_{2r,i}(\u)\}].
\end{eqnarray}
By Assumption (A.1), $$\max_{1\leq k\leq K}\max_{1\leq i\leq n_{k}} \|\x_{ik}\| = o(\max(n_{1}^{1/2}, \cdots, n_{K}^{1/2})) = o(n^{1/2}),$$ we can get
\begin{eqnarray}
&&\frac{r}{n}\sum_{k=1}^K\sum_{i=1}^{n_{k}} E(A_{k,i}(\u))
\nonumber\\
&=& \frac{r}{n}\sum_{k=1}^K\sum_{i=1}^{n_{k}} \sum_{m=1}^M \int_0^{\u^\textsf{T}\tilde{\x}_{ik,m}/\sqrt{r}} \{F(b_{0m}+s)-F(b_{0m})\}ds
\nonumber\\
&=&\frac{\sqrt{r}}{n} \sum_{k=1}^K  \sum_{i=1}^{n_{k}} \sum_{m=1}^M \int_0^{\u^{T}\tilde{\x}_{ik,m}}\{F(b_{0m}+t/\sqrt{r})-F(b_{0m})\}dt
\nonumber\\
&=&\frac{1}{2} \u^\textsf{T}\left( \frac{1}{n}\sum_{k=1}^K\sum_{i=1}^{n_{k}} \sum_{m=1}^M f(b_{0m})\tilde{\x}_{ik,m}\tilde{\x}_{ik,m}^\textsf{T}\right)\u +o(1)
\nonumber\\
&=& \frac{1}{2} \u^\textsf{T}\E\u +o(1).
\label{EA2n}
\end{eqnarray}
Furthermore, we have $$E\left\{\frac{r}{n}\sum_{k=1}^K\sum_{i=1}^{n_{k}} \biggr( A_{k,i}(\u)-E\{A_{k,i}(\u)\}\biggr)\right\}=0,$$
and
\begin{eqnarray}\label{VA2n}
var\biggr(\frac{r}{n}\sum_{k=1}^K\sum_{i=1}^{n_{k}} \left[ A_{k,i}(\u)-E\{A_{k,i}(\u)\}\right]\biggr)
\leq \frac{r^{2}}{ n^2} \sum_{k=1}^K \sum_{i=1}^{n_{k}} E\{A_{k,i}^2(\u)\}.
\end{eqnarray}
Since $A_{k,i}(\u)$ is nonnegative, it is easy to obtain
\begin{eqnarray}
A_{k,i}(\u)&\leq & \bigg\lvert \sum_{m=1}^M \int_0^{\u^\textsf{T}\tilde{\x}_{ik,m}/\sqrt{r}} \{I(\varepsilon_{ik}\leq b_{0m}+s)-I(\varepsilon_{ik}\leq b_{0m})\}ds \bigg\lvert
 \nonumber\\
  &\leq & \sum_{m=1}^M \int_0^{\u^\textsf{T}\tilde{\x}_{ik,m}/\sqrt{r}} \bigg\lvert\{I(\varepsilon_{ik}\leq b_{0m}+s)-I(\varepsilon_{ik}\leq b_{0m})\} \bigg\lvert ds
 \nonumber\\
   &\leq & \frac{1}{\sqrt{r}} \sum_{m=1}^M \mid\u^\textsf{T} \tilde{\x}_{ik,m}\mid.
   \label{boundA2nik}
\end{eqnarray}
By Assumption (A.1), $$\max_{1\leq k\leq K}\max_{1\leq i\leq n_{k}} \|\x_{ik}\| = o(\max(n_{1}^{1/2}, \cdots, n_{K}^{1/2})) = o(n^{1/2}),$$ together with (\ref{VA2n}) and (\ref{boundA2nik}), we get
\begin{eqnarray}\label{covVA2n}
&&var\biggr(\frac{r}{ n}\sum_{k=1}^K\sum_{i=1}^{n_{k}} \left[ A_{k,i}(\u)-E\{A_{k,i}(\u)\}\right]\biggr)
\nonumber\\
&\leq&\left\{M\frac{\|\u\|}{\sqrt{n}}(1+\max_{1\leq k\leq K}\max_{1\leq i \leq n_{k}}\|\x_{ik}\|) \right\}\sum_{k=1}^K\frac{r^{3/2}}{ n^{3/2}}\sum_{i=1}^{n_{k}} E\{ A_{k,i}(\u)\}
\nonumber\\
&=& o(1).
\end{eqnarray}
Combining the Chebyshev's inequality, it follows from (\ref{EA2nk}), (\ref{EA2n}) and (\ref{covVA2n}) that
\begin{eqnarray}\label{A2nuEu}
E\left\{A^*_{2r}(\u)\mid D_n \right\}=\frac{1}{2} \u^\textsf{T}\E\u+o_p(1).
\end{eqnarray}

Next, we derive the conditional variance of $A^*_{2r}(\u)$, i.e., $var \left\{A^*_{2r}(\u)\mid D_n \right\}$. Observing that $A^*_{k,i}(\u), i=1,\cdots, r_{k}$ are independent and identically distributed when given $D_n$,
\begin{eqnarray}\label{VarConA2nk}
var\left\{ A^*_{2r}(\u) \mid D_n\right\}
&=& \sum_{k=1}^K \frac{r^2}{( r_k n)^2} \sum_{i=1}^{r_{k}} var\biggr\{ \frac{A^*_{k,i}(\u)}{\pi^*_{ik}} \biggr\lvert D_n\biggr\}
\nonumber\\
&\leq &\sum_{k=1}^K\frac{r^2 r_k}{r_k^2 n^2}E\biggr[\biggr\{\frac{A^*_{k,i}(\u)}{\pi^*_{ik}}\biggr\}^2 \biggr\lvert D_n\biggr].
\end{eqnarray}
By (\ref{boundA2nik}), the right hand of (\ref{VarConA2nk}) satisfies
\begin{eqnarray}
 &&\sum_{k=1}^K\frac{r^2 r_k}{r_k^2 n^2}\sum_{i=1}^{n_{k}}   \frac{A^2_{k,i}(\u)}{\pi_{ik}}
 \nonumber\\
&\leq& \frac{r^2 }{ n^2}\sum_{k=1}^K\sum_{i=1}^{n_{k}}  A_{k,i}(\u)\biggr(\frac{1}{\sqrt{r}} \sum_{m=1}^M \frac{\mid\u^\textsf{T} \tilde{\x}_{ik,m}\mid}{r_k \pi_{ik}}\biggr)
\nonumber\\
&\leq& \biggr(\frac{r^{1/2} }{ n}M\|\u\|\max_{1\leq k \leq K} \max_{1\leq i \leq n_{k}}\frac{\|\x_{ik}\|+1}{r_k\pi_{ik}}\biggr)\frac{r}{ n}\sum_{k=1}^K \sum_{i=1}^{n_{k}} A_{k,i}(\u).
\label{VarConA2nk2}
\end{eqnarray}
Together with (\ref{EA2n}), (\ref{VarConA2nk2}) and Assumption (A.2), we have
\begin{eqnarray}\label{varasta2n}
var\biggr\{ A^*_{2r}(\u) \mid D_n\biggr\}=o_p(1).
\end{eqnarray}
Together with (\ref{boundA2nik}), (\ref{varasta2n}) and Chebyshev's inequality, we can obtain
\begin{eqnarray}\label{asta2npdn}
  A^*_{2r}(\u)&=&\frac{1}{2}\u^\textsf{T}\E\u+o_{p\mid D_n}(1),
\end{eqnarray}
Here $o_{p\mid D_n}(1)$ means if $\a=o_{p\mid D_n}(1)$, then $a$ converges to 0 in conditional probability given $D_n$ in probability,
in other words, for any $\delta>0$, $P(\mid \a\mid>\delta\mid D_n) \prow 0$ as $n\rightarrow+\infty$. Since $0\leq P(\mid \a\mid>\delta\mid D_n) \leq 1$, then it converges to 0 in probability if and only
$P(\mid \a\mid>\delta)= E\{P(\mid \a\mid>\delta\mid D_n)\} \rightarrow 0$. Thus, $\a=o_{p\mid D_n}(1)$ is equivalent to $\a=o_{p}(1)$.

It follows from (\ref{Andecom0}) and (\ref{asta2npdn}) that
\begin{eqnarray*}
A_{2r}^*(\u) &=&   \u^\textsf{T} \Z_{r}^*+  \frac{1}{2} \u^\textsf{T}\E\u+o_{p}(1).
\end{eqnarray*}
Since $A_{2r}^*(\u)$ is a convex function, we have
\begin{eqnarray*}
\sqrt{r} (\tilde{\vtheta}_s- \vtheta_0) =  - \E_n^{-1} \Z_{r}^*+o_p(1).
\end{eqnarray*}
Based on the above results, we can prove that
\begin{eqnarray*}
 \{\E_n^{-1}\V_\pi \E_n^{-1}\}^{-1/2} \sqrt{r} (\tilde{\vtheta}_s-\vtheta_0) &=&-\{\E_n^{-1}\V_\pi \E_n^{-1}\}^{-1/2} \E^{-1}_n \Z_{r}^*+o_p(1).
\end{eqnarray*}
By Slutsky's Theorem, for any $a\in \mathbf{R}^{p+M}$, from (\ref{Zndistri}) we have that
\begin{eqnarray}\label{pPhi}
P[ \{\E_n^{-1}\V_\pi \E_n^{-1}\}^{-1/2} \sqrt{r} (\tilde{\vtheta}_s-\vtheta_0)\leq \a\mid D_n] \prow \Phi_{p+M}(\a),
\end{eqnarray}
where $\Phi_{p+M}(\a)$ denotes the standard $p+M$ dimensional multivariate normal distribution function.  And the conditional probability in (\ref{pPhi}) is a bounded random variable, then convergence in probability to a constant implies convergence in the mean.
Therefore, for any $\a\in R^{p+M}$,
\begin{eqnarray*}
&&P[ \{\E_n^{-1}\V_\pi \E_n^{-1}\}^{-1/2} \sqrt{r} (\tilde{\vtheta}_s-\vtheta_0)\leq \a]\\
&=&E(P[ \{\E_n^{-1}\V_\pi \E_n^{-1}\}^{-1/2} \sqrt{r} (\tilde{\vtheta}_s-\vtheta_0)\leq \a\mid D_n])\\
&\rightarrow& \Phi_{p+M}(\a).
\end{eqnarray*}
We complete the proof of Theorem 1.

\hspace*{\fill} \\  
\noindent \textbf{Proof the Theorem 2 }

\noindent
We can prove that
\begin{eqnarray*}
tr(\V_\pi) &=& \frac{1}{n^{2}}\sum_{k=1}^{K}\frac{r}{r_{k}} \sum_{i=1}^{n_{k}}\frac{1}{\pi_{ik}}tr \left(\left[\sum_{m=1}^{M}\{I(\varepsilon_{ik} < b_{0m})-\tau_{m}\}\tilde{\x}_{ik,m}\right]^{\otimes2}\right)
 \nonumber\\
 &=&  \frac{1}{n^{2}}\sum_{k=1}^K\frac{r}{r_{k}}\biggr(\sum_{i=1}^{n_{k}} \pi_{ik} \biggr) \left(\sum_{i=1}^{n_{k}}\frac{1}{\pi_{ik}}\biggr\| \sum_{m=1}^M [I(\varepsilon_{ik}<b_{0m})-\tau_m]\tilde{\x}_{ik,m}\biggr\|^2\right)
 \nonumber\\
 &\geq & \frac{1}{n^{2}}\sum_{k=1}^K\frac{r}{r_{k}} \left(\sum_{i=1}^{n_{k}} \biggr\| \sum_{m=1}^M \{I(\varepsilon_{ik}<b_{0m})-\tau_m\}\tilde{\x}_{ik,m}\biggr\|^2\right)
 \nonumber\\
 &=&  \frac{1}{n^{2}}\left(\sum_{k=1}^K r_{k}\right)\sum_{k=1}^K\frac{1}{r_{k}}\left(\sum_{i=1}^{n_{k}}\biggr\| \sum_{m=1}^M [I(\varepsilon_{ik}<b_{0m})-\tau_m]\tilde{\x}_{ik,m}\biggr\|^2\right)
 \nonumber\\
 &\geq & \frac{1}{n^{2}} \sum_{k=1}^K\sum_{i=1}^{n_{k}} \biggr\| \sum_{m=1}^M \{I(\varepsilon_{ik}<b_{0m})-\tau_m\}\tilde{\x}_{ik,m}\biggr\|^2,
\end{eqnarray*}
with Cauchy-Schwarz inequality and the equality in it holds if and only if when $\pi_{ik} \propto \| \sum_{m=1}^M [I(\varepsilon_{ik}<b_{0m})-\tau_m] \tilde{\x}_{ik,m}\|$ and $r_{k} \propto \sum_{i=1}^{n_{k}}\|\sum_{m=1}^M [I(\varepsilon_{ik}<b_{0m})-\tau_m] \tilde{\x}_{ik,m}\|$, respectively. We complete the proof of Theorem 2.

\end{appendices}

\newpage
\makeatletter
\renewcommand\@biblabel[1]{}
\renewenvironment{thebibliography}[1]
     {\section*{\refname}%
      \@mkboth{\MakeUppercase\refname}{\MakeUppercase\refname}%
      \list{\@biblabel{\@arabic\c@enumiv}}%
           {\settowidth\labelwidth{\@biblabel{#1}}%
            \leftmargin\labelwidth
            \advance\leftmargin\labelsep
\advance\leftmargin by 2em%
\itemindent -2em%
            \@openbib@code
            \usecounter{enumiv}%
            \let\p@enumiv\@empty
            \renewcommand\theenumiv{\@arabic\c@enumiv}}%
      \sloppy
      \clubpenalty4000
      \@clubpenalty \clubpenalty
      \widowpenalty4000%
      \sfcode`\.\@m}
     {\def\@noitemerr
       {\@latex@warning{Empty `thebibliography' environment}}%
      \endlist}
\makeatother

\newpage
\begin{table}\begin{center}\small
\title{\small Table 1:   The proposed subsample estimate of $\beta_{1}$ with  $n = 10^{6}$ in Case I.} \\
\label{tab:1}       
\begin{tabular}{ccc ccc ccc ccc}
 \hline
                              &     &$K = 5$ &   &$K = 10$ &   \\
\cmidrule(lr){3-4}  \cmidrule(lr){5-6}
                Error        & $r$ & Bias & SD &  Bias & SD    \\
\hline
                              &200  &~0.0006  &0.0769 &~0.0010  &0.0737 \\
                              &400  &-0.0009  &0.0554 &-0.0008  &0.0531   \\
 $N(0,1)$                     &600  &~0.0025  &0.0425 &~0.0008  &0.0423   \\
                              &800  &~0.0009  &0.0379 &~0.0004  &0.0388  \\
                              &1000 &~0.0004  &0.0348 &-0.0014  &0.0338	  \\
\hline
                              &200  &~0.0023  &0.1405 &~0.0049  &0.1336     \\
                              &400  &-0.0023  &0.0970 &~0.0006  &0.0934  \\
$mixNormal$                   &600  &-0.0033  &0.0797 &-0.0004  &0.0822 \\
                              &800  &~0.0028  &0.0688 &-0.0019  &0.0707 \\
                              &1000 &-0.0002  &0.0600 &-0.0033  &0.0621 \\
\hline
                              &200  &-0.0021  &0.0961 &~0.0009	&0.0914\\
                              &400  &~0.0006  &0.0665 &-0.0004  &0.0645 \\
$ t(3)$                       &600 &-0.0015  &0.0552 &-0.0002  &0.0505  \\
                              &800  &-0.0003  &0.0477 &~0.0005	&0.0462\\
                              &1000 &~0.0024  &0.0415 &~0.0013  &0.0423 \\
\hline
                              &200  &-0.0108  &0.1312 &~0.0070  &0.1373 \\
                              &400  &~0.0040  &0.0959 &~0.0003  &0.0954  \\
$Cauchy$                      &600  &~0.0023  &0.0793 &-0.0008  &0.0778  \\
                              &800  &~0.0011  &0.0700 &-0.0005  &0.0674 \\
                              &1000 &-0.0014  &0.0612 &-0.0018	 &0.0637   \\
\hline
\end{tabular}\end{center}
\end{table}

\begin{table}\begin{center}\small
\title{Table 2: The proposed subsample estimate of $\beta_{1}$ for   Case IV and $\varepsilon \sim N(0,1)$.} \\
\label{tab:5}       
\begin{tabular}{ccc ccc ccc ccc}
\hline
      & $n = 10^{6}$ & & $n = 10^{7}$ &  &     \\
\cmidrule(lr){2-3}  \cmidrule(lr){4-5}
 $r$  & Bias & SD   & Bias & SD      \\
\hline
 200  &~0.0004 & 0.0551  &~0.0005 & 0.0555    \\
 400  &-0.0003 & 0.0394  &~0.0003 & 0.0392   \\
 600  &~0.0002 & 0.0313  &-0.0020 & 0.0312     \\
 800  &~0.0012 & 0.0273  &-0.0005 & 0.0267 \\
 1000 &~0.0012 & 0.0242  &-0.0011 & 0.0256   \\
\hline
\end{tabular}\end{center}
\end{table}

\begin{table}\begin{center}\small
\title{\small Table 3:   The CPU time for Case I and $\varepsilon \sim N(0,1)$ with $K = 5$, $n = 10^{6}$ (seconds)} \\
\label{tab:2}       
\begin{tabular}{ccc ccc ccc ccc}
 \hline
           & $r$ &     &      &     &         \\
\cmidrule(lr){2-6}
$Methods$  & 200     & 400   & 600   & 800   & 1000    \\
\hline
 Uniform   & 0.077   &0.098  &0.145  &0.170  &0.217   \\
 Proposed  & 0.446   &0.494  &0.552  &0.615  &0.689  \\
 Full data & 421.03  &       &       &       &      \\
\hline
\end{tabular}\end{center}
\end{table}

\begin{table}\begin{center}\small
\title{\small Table 4:   The CPU time for  Case I and $\varepsilon \sim N(0,1)$ with $r=1000,K = 5$ and $p=30$ (seconds)} \\
\label{tab:3}       
\begin{tabular}{ccc ccc ccc ccc}
 \hline
           & $n$ &     &     &     \\
\cmidrule(lr){2-5}
$Methods$  & $10^{4}$ & $10^{5}$ & $10^{6}$ & $10^{7}$     \\
\hline
 Uniform   & 0.411    &	0.417    & 0.447    & 0.490   \\
 Proposed  & 0.586	  & 0.620	 & 0.922    & 5.393    \\
 Full data & 4.43     &	61.60    & 676.08   & 4667.22   \\
\hline
\end{tabular}\end{center}
\end{table}

\begin{table}\begin{center}\small
\title{\small Table 5:   The CPs and the average lengths (in parenthesis) of the confident interval of $\beta_{1}$ with $n = 10^{6}$, $r = 1000$ and $K = 5$.} \\
\label{tab:6}       
\begin{tabular}{ccc ccc ccc ccc}
 \hline
       Error              & $B$& Case I &  Case II  &   Case III  &  Case IV       \\
\hline
                                   &20  & 0.930(0.030)  &  0.948(0.034) &   0.932(0.014) &  0.920(0.021) \\
                                   &40  & 0.928(0.021)  &  0.924(0.024) &   0.936(0.010) &  0.954(0.015) \\
 $N(0,1)$    &60  & 0.952(0.018)  &  0.942(0.020) &   0.942(0.009) &  0.944(0.013)  \\
                                   &80  & 0.918(0.015)  &  0.934(0.017) &   0.926(0.008) &  0.914(0.011) \\
                                   &100 & 0.936(0.014)  &  0.934(0.016) &   0.930(0.007) &  0.916(0.010) \\
\hline
                                   &20  &  0.926(0.054) &   0.920(0.060) &   0.938(0.026) &  0.930(0.038)  \\
                                   &40  &  0.932(0.038) &   0.934(0.044) &   0.922(0.019) &  0.954(0.027)  \\
$mixNormal$  &60  &  0.924(0.031) &   0.936(0.036) &   0.930(0.015) &  0.934(0.023)  \\
                                   &80  &  0.928(0.027) &   0.928(0.031) &   0.934(0.014) &  0.946(0.020)  \\
                                   &100 &  0.930(0.025) &   0.934(0.028) &   0.932(0.012) &  0.948(0.018)  \\
\hline
                                   &20  &  0.940(0.037) &   0.940(0.041) &   0.928(0.018) &  0.954(0.026)  \\
                                   &40  &  0.944(0.026) &   0.960(0.030) &   0.946(0.013) &  0.916(0.019)  \\
$t(3)$       &60  &  0.946(0.022) &   0.968(0.025) &   0.936(0.010) &  0.936(0.016)  \\
                                   &80  &  0.940(0.019) &   0.944(0.021) &   0.946(0.009) &  0.940(0.013)  \\
                                   &100 &  0.948(0.017) &   0.944(0.019) &   0.934(0.008) &  0.914(0.012)  \\
\hline
                                   &20  &  0.932(0.053) &   0.944(0.060) &   0.918(0.026) &  0.936(0.038)  \\
                                   &40  &  0.926(0.037) &   0.932(0.043) &   0.922(0.018) &  0.944(0.027)  \\
$ Cauchy$     &60  &  0.924(0.031) &   0.942(0.036) &   0.930(0.015) &  0.926(0.022)  \\
                                   &80  &  0.938(0.027) &   0.946(0.031) &   0.934(0.013) &  0.924(0.020)  \\
                                   &100 &  0.942(0.024) &   0.952(0.028) &   0.926(0.012) &  0.928(0.018)  \\
\hline
\end{tabular}\end{center}
\end{table}

\begin{table}
\begin{center}\small
\title{\small Table 6:   The number of yearly data and allocation sizes $(r = 1000)$} \\
\begin{tabular}{ccc ccc ccc ccc}
\hline
   Years & $n_{k}$ & $r_{k}$  & Years & $n_{k}$ & $r_{k}$   \\
\hline
  1987 &  1,287,333  &  11 &  1998 &  5,227,051   &  45   \\
  1988 &  5,126,498  &  47 &  1999 &  5,360,018   &  45  \\
  1989 &  4,925,482  &  45 &  2000 &  5,481,303   &  45  \\
  1990 &  5,110,527  &  46 &  2001 &  4,873,031   &  42   \\
  1991 &  4,995,005  &  46 &  2002 &  5,093,462   &  45    \\
  1992 &  5,020,651  &  47 &  2003 &  6,375,689   &  56     \\
  1993 &  4,993,587  &  46 &  2004 &  6,987,729   &  59     \\
  1994 &  5,078,411  &  46 &  2005 &  6,992,838   &  58     \\
  1995 &  5,219,140  &  46 &  2006 &  7,003,802   &  57     \\
  1996 &  5,209,326  &  44 &  2007 &  7,275,288   &  58    \\
  1997 &  5,301,999  &  47 &  2008 &  2,319,121   &  19    \\
\hline

\end{tabular}\end{center}
\label{realclean}
\end{table}

\begin{table}\begin{center}\small
\title{\small Table 7:   The estimator and the length of confident interval for $\hat{\vbeta}_L$ with different  $r$ and $B$  for the airline data. } \\
\label{tab:8}       
\begin{tabular}{ccc ccc ccc ccc}
\hline
    & &   B& &         \\
\cmidrule(lr){3-4}
 r   & & 40   & 100    \\
\hline
200 &$\beta_1$ &-0.0524 (-0.0675,-0.0373)   &-0.0458  (-0.0545,-0.0370)     \\
    &$\beta_2$ &0.9232  (0.9164,  0.9299)  &0.9183  (0.9142,0.9225)    \\
    &$\beta_3$ &-0.0242 (-0.0320, -0.0164)  &-0.0221  (-0.0261,-0.0181)     \\

600 &$\beta_1$ &-0.0450 (-0.0539,-0.0361)  &-0.0479  (-0.0537,-0.0421)     \\
    &$\beta_2$ &0.9172  (0.9127,0.9217)  &0.9203  (0.9179,0.9227)   \\
    &$\beta_3$ &-0.0268   (-0.0309,-0.0228)  &-0.0264  (-0.0288,-0.0240)     \\

1000&$\beta_1$ &-0.0446   (-0.0509,-0.0383)  &-0.0404  (-0.0445,-0.0363)    \\
    &$\beta_2$ &0.9192   (0.9163,0.9220)  &0.9205  (0.9184,0.9226)     \\
    &$\beta_3$ &-0.0238   (-0.0269,-0.0208)  &-0.0277  (-0.0297,-0.0257)    \\
\hline
\end{tabular}\end{center}
\end{table}

\begin{figure}
  \includegraphics [width=12cm,height=12cm]{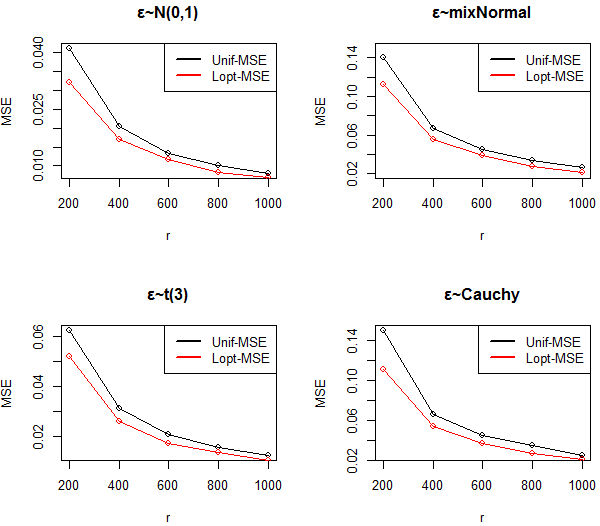}\\
  \caption{The MSEs for different subsampling methods with $K = 5$ and $n = 10^{6}$ (Case 1).}
  \label{f1}
\end{figure}

\begin{figure}
  \includegraphics [width=12cm,height=12cm]{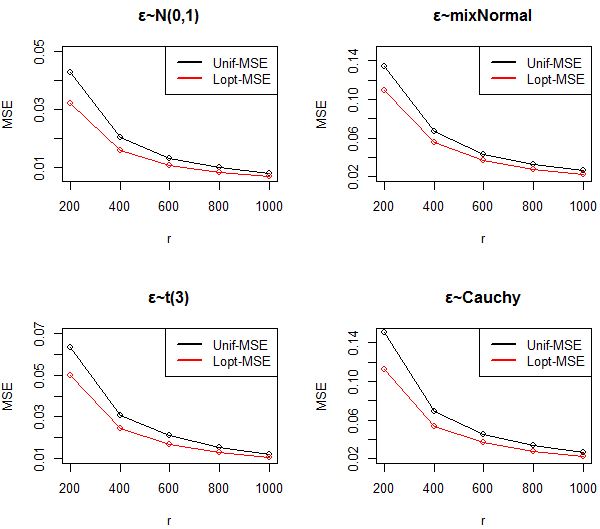}\\
  \caption{The MSEs for different subsampling methods with $K = 10$ and $n = 10^{6}$(Case 1).}
  \label{f2}
\end{figure}

\begin{figure}
  \includegraphics [width=12cm,height=6cm]{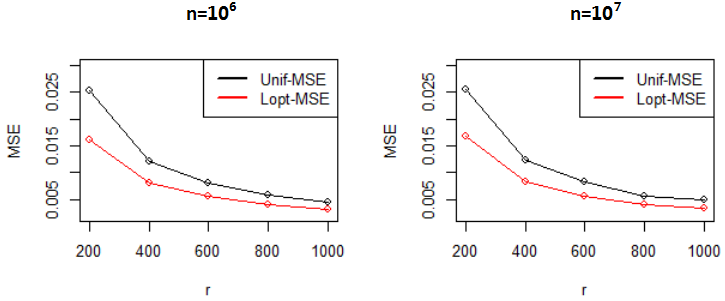}\\
  \caption{The MSEs for different subsampling methods with $\varepsilon \sim N(0,1)$(Case IV). }
  \label{f3}
\end{figure}

\begin{figure}
  \includegraphics [width=12cm,height=12cm]{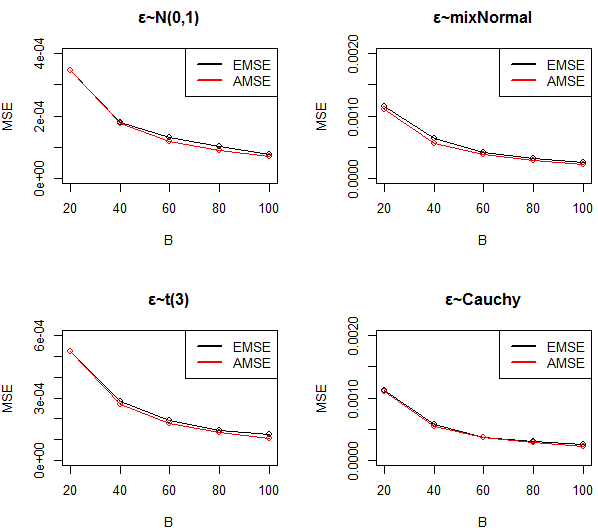}\\
  \caption{The EMSEs and AMSEs of $\hat{\vtheta}_L$ with different values of $B$ and $r =1000$ (Case 1). }
  \label{f4}
\end{figure}

\begin{figure}
  \includegraphics [width=12cm,height=12cm]{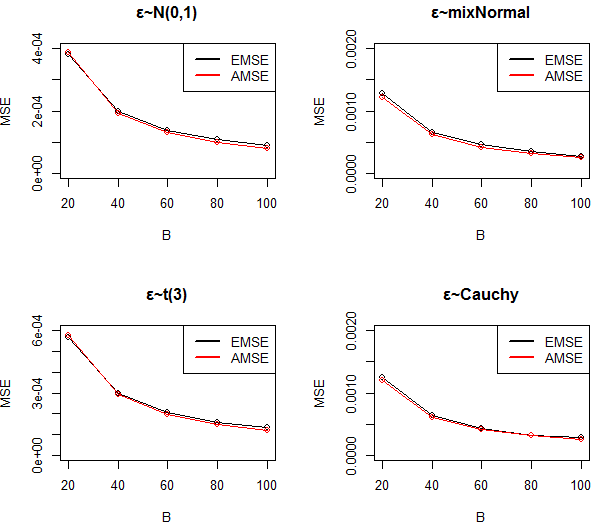}\\
  \caption{The EMSEs and AMSEs of $\hat{\vtheta}_L$ with different values of $B$ and $r =1000$ (Case II).}
  \label{f5}
\end{figure}

\begin{figure}
  \includegraphics [width=12cm,height=12cm]{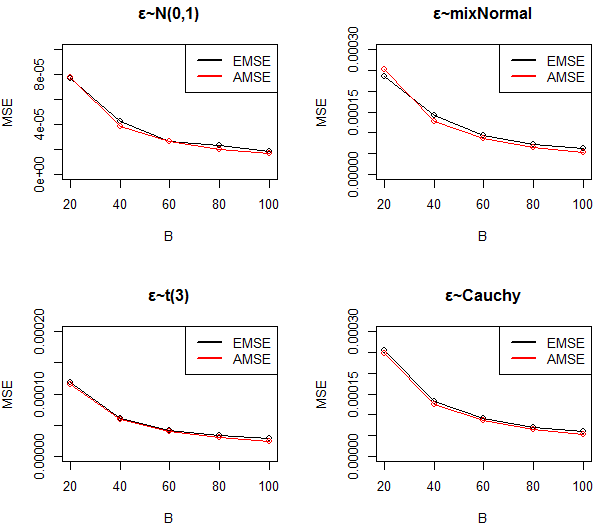}\\
  \caption{The EMSEs and AMSEs of $\hat{\vtheta}_L$ with different values of $B$ and $r =1000$ (Case III). }
  \label{f6}
\end{figure}

\begin{figure}
  \includegraphics [width=12cm,height=12cm]{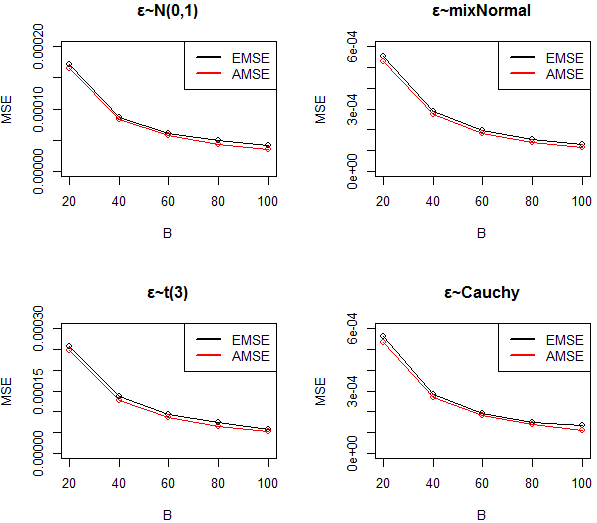}\\
  \caption{The EMSEs and AMSEs of $\hat{\vtheta}_{L}$ with different values of $B$ and $r =1000$ (Case IV).}
  \label{f7}
\end{figure}

\begin{figure}
  \centering
  \includegraphics [width=6cm,height=6cm]{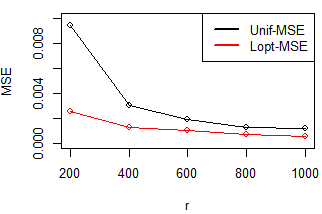}\\
  \caption{The results of MSEs for the airline data.}
  \label{f8}
\end{figure}


\bibliography{sn-bibliography}


\end{document}